\documentclass[epj]{svjour}
\usepackage{graphics}
\usepackage{graphicx}
\begin{document}
\title{Polarization of Hyperons in Elementary Photoproduction}
\author{Reinhard Schumacher\inst{1} 
}                     
\mail{schumacher@cmu.edu}          
\institute{Department of Physics, Carnegie Mellon University, Pittsburgh, PA 15213, USA}
%
\date{Received: HYP2006 Proceedings Article, 12-14-2006 }
%
\abstract{ Recent measurements using the CLAS detector at Jefferson
Lab of the reactions $\vec\gamma + p \to K^+ + \vec\Lambda$ and
$\vec\gamma + p \to K^+ + \vec\Sigma^0$ have been used to extract the
spin transfer coefficients $C_x$ and $C_z$ for the first time.  These
observables quantify the degree of the photon circular polarization
that is transferred to the recoiling hyperons in the scattering plane.
The unexpected result is that $\Lambda$ hyperons are produced ``100\%
polarized'' as seen when combining $C_x$ and $C_z$ with the induced
transverse polarization, $P$.  Furthermore, $C_x$ and $C_z$ seem to be
linearly related.  This paper discusses the experimental results and
offers a hypothesis which can explain these observations.  We show how
the produced strange quark can be subject to a pure spin-orbit type of
interaction which preserves its state of polarization throughout the
hadronization process.
} 
\maketitle
\section{Introduction}
\label{intro}
At the HYP2006 Conference in Mainz in October 2006 I presented a talk
entitled ``Experiments with Strangeness in Hall B at Jefferson Lab.''
There were two topics: first, the measurement of the spin transfer
coefficients $C_x$ and $C_z$ in $K^+Y$ production off the proton using
real photons~\cite{bradford}, and second, the measurement of four
separated cross section components in $K^+Y$
electroproduction~\cite{ambrozewicz}.  Both topics are the subjects of
long papers that have since been submitted for publication, as cited,
and essentially all points made in the talk are covered in those two
papers.

Rather than repeat that discussion, this paper will provide further
details about the $C_x$ and $C_z$ spin transfer work that were partly
mentioned in the talk, but not covered in Ref.~\cite{bradford}.  Two
phenomenological puzzles were presented in the talk and the paper
regarding the new polarization observables.  First, the magnitude of
the $\Lambda$ polarization vector, $|\vec P_\Lambda|$, comprised of
three measured orthogonal components, is unity at all production
angles and for all center of mass (c.m.) energies $W$.  For a fully
polarized photon beam, $\vec P_\Lambda$ is equivalent to a quantity we
introduce called $\vec R$, defined as $\vec R = C_x \hat x + P \hat y
+ C_z \hat z$. The component $P$ is the induced or transverse
polarization, using the notation common in the literature for this
quantity.  We find $|\vec R| = \sqrt{C_x^2 + C_z^2 + P^2} = 1$ to very
good precision.  Second, there appears to be a simple linear
relationship between the two spin transfer coefficients, to wit, $C_z
= C_x + 1$.  Both of these observation may be considered quite
unexpected since there is no {\it a priori} reason for the $\Lambda$
hyperon polarization to be $100\%$, nor is there an obvious
relationship among the production amplitudes discussed in the
literature to lead to this result.  Indeed, no present theoretical
models incorporates these new pieces of phenomenology.  In this paper
I present a somewhat heuristic model, or hypothesis, which can explain
these findings and which may be a foundation for additional
theoretical work.

\section{Methods, Formalism and Results}
\label{sec:formal}
\subsection{Measurement Method}
\label{sec:methods}

An energy-tagged real photon beam was created in the Hall-B beam line
at Jefferson Lab between energies corresponding to the hyperon
production threshold near $W=1.679$ GeV and 2.454 GeV.  The electron
beams that created the photons via bremsstrahlung were longitudinally
polarized at about $65\%$, for beam energies at 2.4 and 2.9 GeV.  This
longitudinal polarization was transferred as circular polarization of
the created photons during in a well-defined way during
bremsstrahlung, with maximum transfer at the endpoints.  An
unpolarized 18 cm long liquid hydrogen target was used. The CLAS
detector was triggered by any single charged-particle track, including
pions, kaons, and protons.  For this analysis, a positive kaon track
and a proton track from hyperon decay were required to be present, but
the $\pi^-$ from the hyperon decay was not used.  Differential cross
sections and the induced recoil polarization $P$ from this experiment
were published previously~\cite{mcnabb}.

\begin{figure}
\includegraphics[scale=0.35,angle=-90.0]{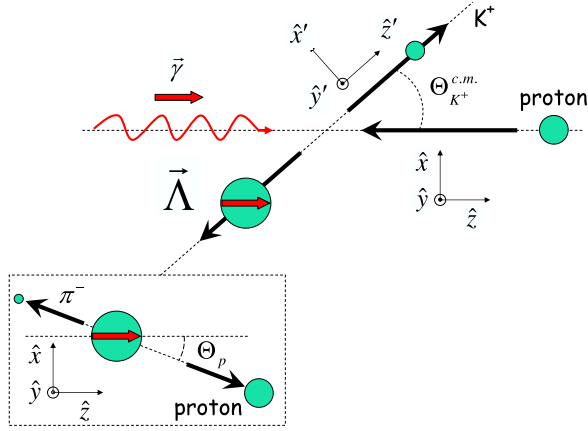}
\caption{ In the overall reaction center of mass, the coordinate
system can be oriented along the outgoing $K^+$ meson
$\{\hat{x}^\prime,\hat{y}^\prime,\hat{z}^\prime\}$ or along the
incident photon direction $\{\hat{x},\hat{y},\hat{z}\}$.  The dotted
box represents the rest frame of the hyperon, and the coordinate
system used for specifying the polarization components. The short
heavy arrows represent polarization vectors.}
\label{fig:axes}       
\end{figure}

Figure~\ref{fig:axes} shows our axis convention.  In the c.m. frame we
adopt the $\{\hat x$, $\hat y$, $\hat z\}$ system wherein the $\hat z$
axis points along the photon direction; it is the most natural one in
which to present these results.  The $\Lambda$ hyperon is produced
polarized, and we can measure the components of this polarization
using the parity-violating weak decay asymmetry of the protons (or
pions) in the rest frame of the $\Lambda$, as illustrated in the
dotted box.  All three components of the polarization can be extracted
by projecting along the relevant axes.  In the specified coordinate
system $i\in\{x,y,z\}$ is one of the three axes.  The decay
distribution, $I_i(\cos \theta_i)$, is given by
\begin{equation}
I_i(\cos \theta_i) = {\textstyle\frac{1}{2}}(1 + \nu \alpha P_{Y i} \cos \theta_i),
\label{eq:decay}
\end{equation}
where $\theta_i$ is the proton polar angle with respect to the given
axis in the hyperon rest frame.  The weak decay asymmetry, $\alpha$,
is taken to be $0.642$.  The factor $\nu$ is a ``dilution'' arising in
the $\Sigma^0$ case due to its radiative decay to a $\Lambda$, and
which is equal to $-1/3$ in the $\Lambda$ rest frame.  A complication
arose for us because we measured the proton angular distribution in
the rest frame of the parent $\Sigma^0$.  This led to a value of $\nu
= -1/3.90$, as discussed in Ref.~\cite{bradford}.  For the
$K^+\Lambda$ analysis $\nu=+1.0$.  Extraction of $P_{Y i}$ follows
from fitting the linear relationship of $I_i(\cos \theta_i)$ vs.~$\cos
\theta_i$.  The polarization of the hyperon in the c.m. frame is the
same as it is in the hyperon rest frame in this experiment: there is
no Wigner rotation when boosting from the hyperon rest frame to the
c.m. frame~\cite{bradford}.

Let $P_\odot$ represent the degree of beam polarization between
$-1.0$ and $+1.0$.  Then the spin-dependent cross section for $K^+Y$
photoproduction can be expressed as~\cite{barker}
\begin{equation}
\rho_Y \frac{d\sigma}{d\Omega_{K^+}} =
\left. {\frac{d\sigma}{d\Omega_{K^+}}}\right| _{unpol.} \!\!\!\!\!\!\!\!\!\!\!
\left\{ 1 + \sigma_y P + 
P_\odot (C_x \sigma_x + C_z \sigma_z ) \right \}.
\label{eq:cross}
\end{equation}
Here $\rho_Y$ is twice the density matrix of the ensemble of recoiling
hyperons $Y$ and is written
\begin{equation}
\rho_Y = (1 + \vec\sigma\cdot \vec P_Y),
\label{eq:density}
\end{equation}
where $\vec \sigma$ are the Pauli spin matrices and $\vec P_Y$ is the
measured polarization of the recoiling hyperons.  In Eq.~\ref{eq:cross}
the spin observables are the induced polarization $P$, and the
polarization transfer coefficients $C_x$ and $C_z$.  

We define our $C_x$ and $C_z$ with signs opposite to the version of
Eq.~\ref{eq:cross} given in Ref.~\cite{barker}.  This makes $C_z$
positive when the $\hat z$ and $\hat z^\prime$ axes coincide at the
forward meson production angle, meaning that positive photon helicity
results in positive hyperon polarization along $\hat z$.

The connection between the measured hyperon recoil polarization vector
$\vec P_Y$ and the spin correlation observables $P$, $C_x$, and $C_z$,
is obtained by taking the expectation value of the spin operator
$\vec\sigma$ with the density matrix $\rho_Y$ via the trace: $\vec P_Y
= Tr(\rho_Y \vec \sigma)$. This leads to the identifications
\begin{eqnarray}
P_{Y x} &=& P_\odot C_x 
\label{eq:obsx} \\
P_{Y y} &=& P 
\label{eq:obsp} \\
P_{Y z} &=& P_\odot C_z. 
\label{eq:obsz}
\end{eqnarray}
The transverse or induced polarization of the hyperon, $P_{Y y}$, is
equivalent to the observable $P$, while the $\hat x$ and $\hat z$
components of the hyperon polarization are proportional to $C_x$ and
$C_z$ via the beam polarization factor $P_\odot$.  Physically, $C_x$
and $C_z$ measure the transfer of circular polarization, or helicity,
of the incident photon on an unpolarized target to the produced
hyperon.

To extract $C_x$ and $C_z$ the beam helicity asymmetry $A_i$ was
accumulated in each bin of proton decay angle with respect to the
$\hat x$ or $\hat z$ axis, and a fit to this asymmetry as a function
of $\cos \theta_i$ was made.  $C_i$ was computed from
\begin{equation}
A(\cos\theta_i) = \frac{N_+ - N_-}{N_+ + N_-} =
\alpha\nu P_\odot C_i \cos\theta_i. 
\label{eq:asym}
\end{equation}
where $N_\pm$ are the helicity-dependent yields in each bin.  The
overall systematic uncertainty for the $K^+\Lambda$ results was $\pm
0.03$ for $\cos \theta^{c.m.}_{K^+} < 0.55$ and $\pm 0.09$ for $\cos
\theta^{c.m.}_{K^+} > 0.55$.  The total global systematic uncertainty
for the $K^+\Sigma^0$ results as $\pm 0.03$ for $\cos
\theta^{c.m.}_{K^+} < 0.35$ and $\pm 0.17$ for $\cos
\theta^{c.m.}_{K^+} > 0.35$.  The systematic uncertainty in $W$ was
$\pm 2$ MeV at the bin centers.  More thorough discussion of the
experimental and analysis details can be found in
Ref.~\cite{bradford}.

\subsection{Experimental Results}
\label{sec:results}
\begin{figure}
\resizebox{0.50\textwidth}{!}{\includegraphics{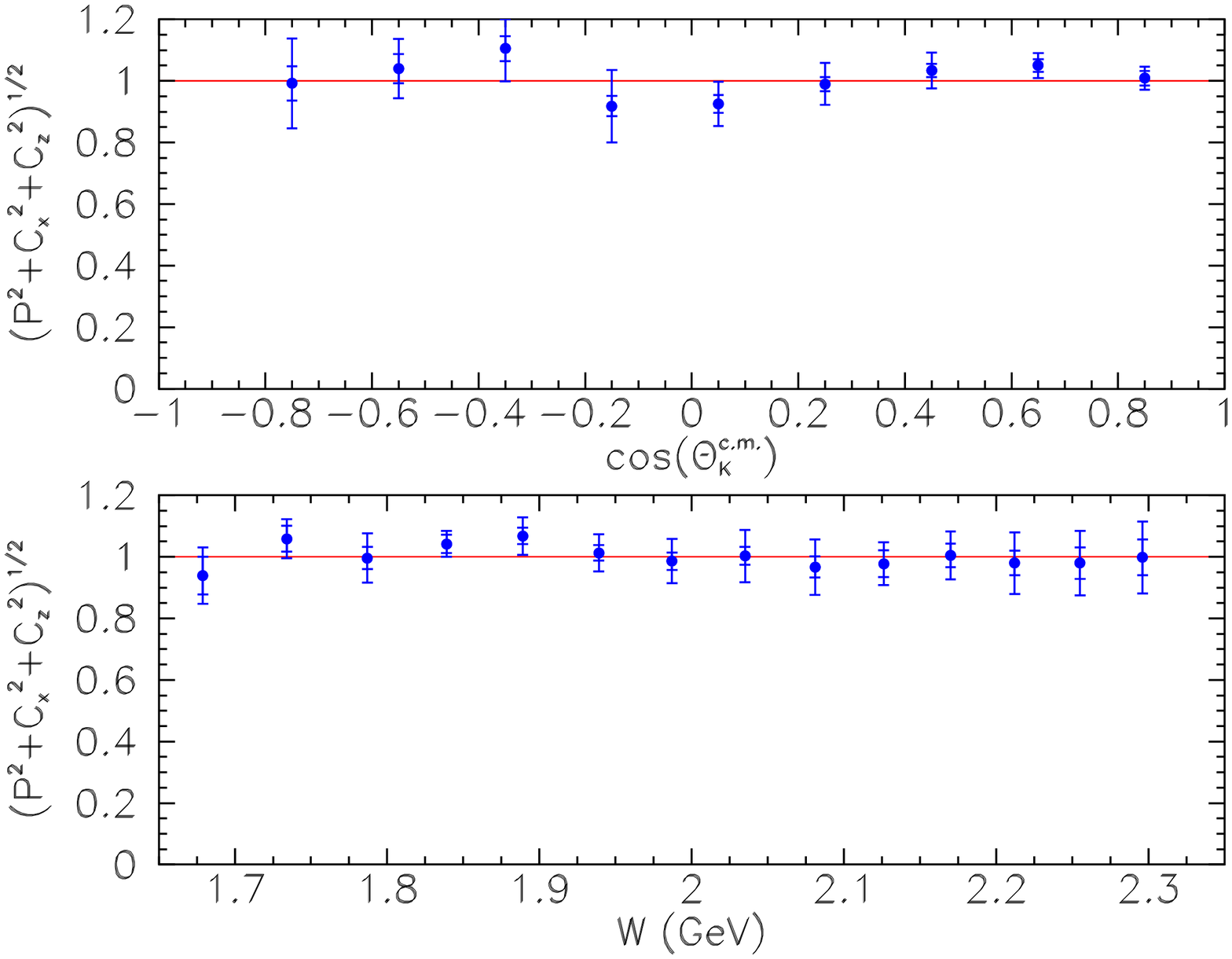}}
\caption{Magnitude of the polarization of the $\Lambda$ hyperon when
  averaging over all measured energies and given as a function of angle
  (top), and alternatively when averaged over all angles given as a function
  of c.m. energy (bottom).  The magnitude, $R_\Lambda$, is consistent
  with unity everywhere. The error bars are discussed in the text.}
\label{fig:lambda}       
\end{figure}

Values of $C_x$ and $C_z$ in their smallest binning of $W$ and
$\cos\theta_{K^+}^{c.m.}$ are presented in Ref.~\cite{bradford}.
Also shown is that the values of 
\begin{equation}
|\vec R_\Lambda| \equiv \sqrt{P^2 + C_x^2 +C_z^2} 
\label{eq:are}
\end{equation}
in all bins are remarkably close to unity for the $\Lambda$.  To
emphasize this more clearly in the present paper, in
Fig.~\ref{fig:lambda} we show the effect of averaging those results
for the $|\vec R_\Lambda|$ across all values of $W$ and showing the
results as a function of kaon production angle (top panel), or
averaging over all angles and showing the results as a function of
$W$.  The points are the weighted mean of the data.  The inner error
bars on each point correspond to the uncertainty on the weighted mean
of the data.  However, taking a weighted mean is strictly appropriate
only if one knows that the set of values to be combined measure the
same physical quantity.  This experiment has discovered that the
values seem to be consistent with unity, but a more fair way of
representing the spread of these values, absent certain knowledge that
they {\it should} be the same, is to use the weighted variance.  The
latter is shown as the outer error bars on the points.  Some data
points lie above unity by about one full error bar, but this is to be
expected based on the analysis method and random error statistics: the
fitted asymmetries were not biased by imposition of the physical limit
at $R=+1.0$.

The unexpected observation is that across all measured angles and
energies the value of $|\vec R_\Lambda|$ is consistent with unity.
Taking the grand weighted mean over the results at all energies and
angles we find
\begin{equation}
\overline{R}_\Lambda = 1.01 \pm 0.01,
\end{equation}
where the uncertainty is that of the weighted mean.  Our systematic
uncertainty is about $\pm 0.03$.  The $\chi^2$ for a fit to the
hypothesis that $|\vec R_\Lambda| = 1$ is 145 for 123 degrees of
freedom, for a reduced chi-square of 1.18, which is a good fit.  One
may therefore conclude that the $\Lambda$ hyperons produced in
$\vec\gamma + p \to K^+ + \vec\Lambda$ with circularly polarized
photons appear 100\% spin polarized.  This result is only ``natural''
at extreme forward and backward angles where the $K^+ \Lambda$ final
state system has no orbital angular momentum available, and all of the
photons' helicity must end up carried by the hyperon.  Since this
situation is not {\it required} by the kinematics of the reaction,
there must be some dynamical origin of this phenomenon, as discussed
below.

Shown in Ref.~\cite{bradford} is that $C_z$ is large and positive over
most of the kinematic range, except at back angles where considerable
``resonance-like'' fluctuations are seen.  The observable $C_x$ has
similar fluctuations as $C_z$ but is typically smaller by one full
unit, meaning that to a good approximation
\begin{equation}
C_z \simeq C_x + 1.
\label{eq:cxczone}
\end{equation}
This is the second unexpected observation about the results of this
experiment.  Taking the weighted mean of the difference $D \equiv
C_z-C_x-1$ over all values of $W$ and kaon angle leads to the value
\begin{equation}
\overline D = 0.054 \pm 0.012
\end{equation}  
In this case the $\chi^2$ for a fit to the hypothesis of
Eq.~\ref{eq:cxczone} is 306 for 159 degrees of freedom, or 1.92 for the
reduced $\chi^2$.  This is a poor fit, so our confidence in the
accuracy of this simple empirical relationship is limited, and
indicates that it needs experimental confirmation.  Nevertheless, we
offer a possible reason for this relationship below.

\begin{figure}
\resizebox{0.50\textwidth}{!}{\includegraphics{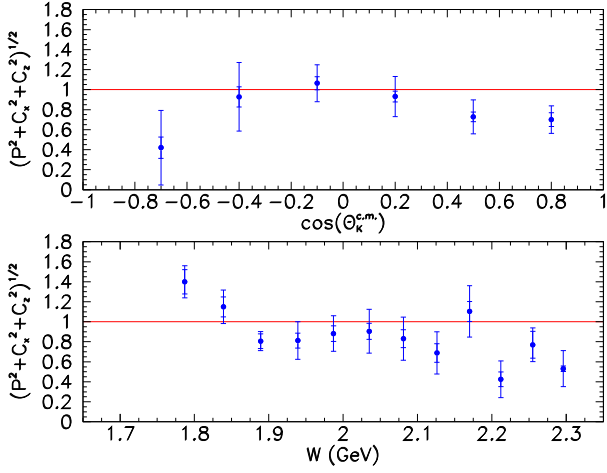}}
\caption{Magnitude of the polarization of the $\Sigma^0$ hyperon when
  averaging over all measured energies and given as a function of
  angle (top), and alternatively when averaged over all angles given
  as a function of c.m. energy (bottom).  The magnitude,
  $R_{\Sigma^0}$, is not statistically consistent with unity.  The
  error bars are discussed in the text.}
\label{fig:sigma0}       
\end{figure}

In the case of the $\Sigma^0$ hyperon the results are less clear
cut.   Figure~\ref{fig:sigma0} shows the same correlations as in the
previous figures.  The reduced statistical precision comes from the
dilution of the $\Sigma^0$ polarization information due to its
radiative decay; its production cross section is, to first order, the
same as that of the $\Lambda$~\cite{bradforddsdo}.  It appears that
the weighted mean values of $R_{\Sigma^0}$ are generally large, but
not consistently close to unity as was the case with the
$\Lambda$.    We found that the angle and energy averaged value is
\begin{equation}
\overline R_{\Sigma^0} = 0.82 \pm .03.
\end{equation}
Thus, this hyperon is not produced ``fully polarized'' from a fully
polarized beam.  In a valence quark picture the $\Sigma^0$ spin is
carried by a combination of the $s$ quark spin and a triplet $ud$
quark spin, unlike in the case of the $\Lambda$ where the $ud$ quarks are in
a spin singlet.  For the following discussion we ignore the $\Sigma^0$
since we can not argue that the strange quark polarization is
manifest as the hyperon polarization without a scale factor.

\section{The Model Hypothesis}
\label{sec:model}

The problem at hand is to deduce why the $\Lambda$ polarization in the
reaction $\gamma p \to K^+ \Lambda$, with fully circularly polarized
photons is ``100\%''.  This is not a feature of any of at least six
highly-developed theoretical models that have been compared with these
experimental results, as shown and discussed in Ref.~\cite{bradford}.

\begin{figure}
\includegraphics[scale=0.40,angle=-90.0]{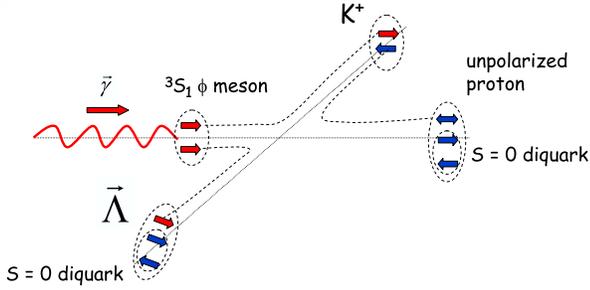}
\caption{Quark-line cartoon to illustrate one possible scenario in
  this hypothesis.  An $\overline s s$ quark pair produced from the
  photon hadronizes such that the $s$ quark in the $\Lambda$ retains
  its full polarization after being ``precessed'' by a
  spin-orbit interaction, while the $\overline s$ quark ends up
  in the spinless kaon.}
\label{fig:quarks}       
\end{figure}

Our {\it ansatz} is that the reaction proceeds via the creation of a
virtual $\overline s s$ quark pair in a $^3S_1$ state. A virtual
$\phi$ meson is created, in a vector dominance picture, that carries
the polarization of the incoming photon, as illustrated in
Fig.~\ref{fig:quarks}.  Alternatively, the $\overline s s$ pair is
created as part of a complex interaction in the gluon field of the
nucleon.  Either way, the key assumption is that the $s$ quark is
produced in a pure spin state.  Next we demand that this polarized
quark survives the hadronization process into the final state
$\Lambda$ in the form of a pure spin state.  We further assume that
the $\Lambda$ spin polarization is a faithful representation of that
of the $s$ quark contained within it.  One can then ask what form the
interaction Hamiltonian may take, such that the quark spin is not
changed in magnitude but only in its orientation.  A possible answer
is given by the theory of two-component spinor dynamics~\cite{merz}.

In keeping with the ansatz, we construe the interaction to be between
a spin $\frac{1}{2}$ quark in the field of a nucleon.  The nucleon also has spin
$\frac{1}{2}$, but it is unpolarized, so we will for this discussion pretend
that it is effectively spinless.  Consider the initial quark spin
state $\chi_0$ to be a linear superposition of eigenstates with
respect to the beam ($\hat z$) axis taken as positive and negative
helicity states $\alpha$ and $\beta$.  A fully polarized quark has
helicity +1 and is in state $\alpha$.

The interaction Hamiltonian we will consider (because it has the
desired property) is that of a spin-orbit interaction between the
quark spin and the orbital angular momentum, $\vec L$, of the quark
with respect to the hardronizing nucleon system:
\begin{equation}
H=\frac{\vec p ^2}{2m} + V(r) + W(r) \vec L \cdot \vec \sigma
\label{eq:ham}
\end{equation}
where $V(r)$ is the spin-independent central potential, $W(r)$ is the
spin-dependent potential, and $\vec \sigma$ are the Pauli spin
matrices that act upon $\chi_0$.  A spin-orbit interaction of the form
given in Eq.~\ref{eq:ham} arises from, for example, a magnetic dipole
(of the quark) interaction with an induced magnetic field (due to the
quark moving in the electric field of the nucleon).  $H$ is a scalar
invariant under rotations and reflections, which is the key property
needed to obtain the desired result.  A rotationally invariant $H$
commutes with $\vec J = \vec L + \vec S$.  An incoming state of given
helicity, $\alpha$ or $\beta$, is in an eigenstate of $J_z = \pm
\frac{1}{2} \hbar$, and the scattering state must have with same
$J_z$.  Reflection through any plane must also leave the scattering
state unaltered. The scattering matrix, $S$, that acts on $\chi_0$,
after also considering these requirements of rotational and reflection
symmetry, has the form
\begin{equation}
S =
\left(
\begin{array}{cc}
g(\theta)              & h(\theta)e^{-i \phi} \\
-h(\theta)e^{i \phi} & g(\theta)
\end{array}
\right)
\label{eq:matrix}
\end{equation}
where $g(\theta)$ is a complex non-spin-flip amplitude, $h(\theta)$
is a complex spin-flip amplitude, and $\phi$ is the azimuthal
scattering angle.  That is, $h(\theta)$ turns $\beta$
into $\alpha$, with some phase factor, while $g(\theta)$ leaves
$\alpha$ as $\alpha$ modified by a phase.  The polar angular
dependence on the production angle, $\theta$, can only be determined
by actually solving the scattering equation.  Using Euler's formula
and the Pauli matrices, this is equivalent to
\begin{equation}
S = g(\theta) 1 + i h(\theta) \hat n \cdot \vec \sigma
\label{eq:scatter}
\end{equation}
where $\hat n = (\vec \gamma \times \hat K^+)/|\vec \gamma \times \hat
K^+|$ is normal to the scattering plane.  Having only two complex
amplitudes is not accurate, since the nucleon actually has spin $\frac{1}{2}$ as
well, leading to four complex amplitudes for pseudo-scalar meson
photoproduction.  We proceed anyway, since the proton spin is not
polarized and the categorization into spin-flipping and non-flipping
amplitudes for the quark itself retains some generality.

The final spin state of the quark, $\chi_f$, is given by $S \chi_0$.
The polarization $\vec P_f$ of the final spin state is given by the
expectation value of $\vec \sigma$, with suitable normalization in the
denominator of the final-state spin vector:
\begin{equation}
\vec P_f = <\vec\sigma> = \frac{\chi_f^\dagger \vec \sigma
  \chi_f}{\chi_f^\dagger \chi_f} =
\frac{\chi_0^\dagger S^\dagger \vec \sigma S \chi_0}{\chi_0^\dagger S^\dagger S \chi_0}
\label{eq:polfinal}
\end{equation}
To evaluate this expression we use the density matrix
formulation of the initial and final spin states as the way to capture
all the phase information in the computation.  We use $\rho_0 = \chi_0
\chi_0^\dagger = \frac{1}{2}(1+\vec P_0\cdot\vec\sigma)$ and $\rho_f =
\chi_f \chi_f^\dagger = \frac{1}{2}(1+\vec P_f\cdot\vec\sigma)$ and
the fundamental formula for evaluating expectations values of an
operator $\vec A$, that is, $<\vec A> = Tr(\rho \vec A)$.
The final state polarization becomes
\begin{equation}
\vec P_f = \frac{Tr(\rho_0 S^\dagger \vec\sigma S)}{Tr(\rho_0 S^\dagger  S)}
\label{eq:polfinal2}
\end{equation}
The denominator in this expression is the differential cross section
$d\sigma/d\Omega$.  Substituting the expressions for $S$ and $\rho_0$,
and bringing to bear the necessary trace identities reduces this to
\begin{eqnarray}
\lefteqn{\vec P_f (g^*g+h^*h + i(g^*h - h^*g)(\vec P_0 \cdot \hat n) } \nonumber \\
  & = & (i(g^*h-h^*g)+2 h^*h \vec P_0 \cdot \hat n)\hat n + \nonumber \\
  &   & (g^*g - h^*h)\vec P_0 + (g^*h + h^*g)(\vec P_0 \times \hat n)
\label{eq:polfinal3}
\end{eqnarray}
For the problem at hand the initial quark polarization is along the
beam line, the $\hat z$ direction, and can be written $\vec P_0 =
(0,0,P_\odot)$.  Now the three terms in Eq.~\ref{eq:polfinal3} are
orthogonal and can be identified with the $\hat y$, $\hat z$, and
$-\hat x$ directions.  Note the sign reversal on $\hat x$ to be consistent
with our coordinate system choices in Fig.~\ref{fig:axes}.  

We thus have four observables for determining the four real parameters
of the two complex amplitudes $g(\theta)$ and $h(\theta)$.
\begin{eqnarray}
\frac{d\sigma}{d\Omega} =& g^*g + h^*h & 
\label{eq:dsdo} \\
P_{fx} =& \frac{g^*h + h^*g}{g^*g + h^*h} P_\odot &\equiv -C_x P_\odot 
\label{eq:pfx} \\
P_{fy} =& i \frac{g^*h - h^*g}{g^*g + h^*h}  &\equiv P 
\label{eq:pfy} \\
P_{fz} =& \frac{g^*g - h^*h}{g^*g + h^*h} P_\odot &\equiv C_z P_\odot
\label{eq:pfz}
\end{eqnarray}

It is easy to check from these equations, having started with a
rotationally symmetric spin-orbit Hamiltonian and written it in terms
of two amplitudes, that the magnitude of the polarization vector is
given by $|\vec R| = 1$ for any $g(\theta)$ and $h(\theta)$ when
$P_\odot = 1$.  That is, a spin-orbit type of interaction preserves
the magnitude of the polarization and only rotates it in some fashion.
No constraints are placed on the forms of $g(\theta)$ and $h(\theta)$.

The components of the amplitudes can be determined from the
experimental results directly.  If we write $g = g_r e^{i\phi_g}$ and
$h = h_r e^{i \phi_h}$ then we find
\begin{eqnarray}
g_r &=& \left (\textstyle\frac{1}{2}\frac{d\sigma}{d\Omega}(1+C_z)\right )^{1/2}, \\
h_r &=& \left (\textstyle\frac{1}{2}\frac{d\sigma}{d\Omega}(1-C_z)\right )^{1/2}, \\
\Delta \phi &=& \tan^{-1} \frac{P}{C_x},
\end{eqnarray}
where $\Delta \phi = \phi_g - \phi_h$ is the phase difference between
$g$ and $h$.  The overall phase is unimportant.

Preservation of the polarization magnitude is a general property of
interactions of the form Eq.~\ref{eq:ham}, as can also be
derived by investigating~\cite{merz} the time dependence of a
polarization vector $\vec P(t)$.  By writing the time-dependent
Schr\"odinger equation for $\chi_0$ and evaluating the time dependence
of the expectation value of $\vec \sigma$ one is led to
\begin{equation}
2\hbar \frac{d\vec P}{d t} = \vec L \times \vec P.
\label{eq:timedep}
\end{equation}
This means that the change in the polarization vector is perpendicular
to the vector itself, so that $\vec P$ ``precesses'' around $\vec L$
in the manner of a magnetic moment precessing around a static magnetic
field.

The second puzzle in the experimental results was the observation that
$C_z \simeq C_x + 1$.  From Eqs.~\ref{eq:pfx} and ~\ref{eq:pfz}, this
observation has the consequence that
\begin{equation}
h^*h = \textstyle\frac{1}{2}(g^*h + h^*g).
\label{eq:hh}
\end{equation}
Again using the polar representation, this expression shows a simple
relationship between the magnitudes of the spin-flip amplitude
$h(\theta)$ and the non spin-flip amplitude $g(\theta)$:
\begin{equation}
h_r = g_r \cos \Delta \phi,
\label{eq:epsilon}
\end{equation}
where the phase difference $\Delta\phi$ could be a function
of $W$ and $\theta$.  It is shown in Ref.~\cite{bradford} that $C_z$
it is large and positive in most regions of energy and angle.  This
means the non spin-flip amplitude $g$ is dominant and the spin-flip
amplitude $h$ is small in most regions of energy and angle.  Therefore
$\cos \Delta \phi$ is a fairly small number and the phase
difference $|\Delta\phi|$ is near $\pi/2$.  

Thus, the second puzzle would find its explanation in a phenomenology
wherein the two interfering amplitudes in this process, spin-flip and
non spin-flip, are everywhere proportional to the cosine of their
phase difference.  The accuracy of this statement hinges on an
experimental result that is only of modest precision, as discussed
above, and underlines the desirability to make further experimental
checks of this interpretation.

\section{Discussion and Conclusion}
\label{sec:conclusion}

We note that this general picture of the reaction process diverges
from the notion that the photon first produces a non-strange $N^*$ or
$\Delta$ baryon, and that this baryon then couples to the $K^+\Lambda$
final state.  The latter approach does not build in the idea that the
spin of the created strange quark is intimately tied to the spin of
the incoming photon.  In the standard picture, the spin couplings are
treated at the hadronic level, not the quark level.  The model
hypothesis discussed here was developed largely to account for such a
connection.

This model hypothesis can make some testable predictions since there
are other observables that have not been measured but ought to have
behavior explained in this picture.  An example is the case when the
photons are linearly polarized and the hyperon recoil polarization is
again measured.  In that case there are the observables $O_x$ and
$O_z$.  The picture suggested here would predict the initial creation
of a transversely polarized $s$ quark as part of a $^3S_1$ pair,
followed by again a spin-orbit like quark-baryon hadronization
interaction that preserves the polarization magnitude.  The induced
polarization $P$ would again be the third orthogonal component,
leading to the prediction
\begin{equation}
O_x^2 + O_z^2 + P^2 = 1
\label{eq:oxoz}
\end{equation}
There are a large number of similar predictions which could be
examined in the light of this hypothesis.

In summary, this paper has presented some details of recent first-time
measurements of the beam-recoil spin-transfer measurements for $K^+Y$
photoproduction on the proton.  Two puzzles that were raised in a
recent talk/paper were discussed.  First, why is the net polarization
of the $\Lambda$ hyperon in the $K^+\Lambda$ final state essentially
$100\%$?  Second, why is the transferred spin polarization in the
$\hat z$ direction, $C_z$, one unit larger than the in-plane
transferred polarization in the $\hat x$ direction, $C_x$?  A model
explanation was presented that is built on the idea that the created
$s$ quark carries the photon polarization as a pure spin state, and
that this quark experiences a spin-orbit type of interaction during
hadronization that allows it to ``precess'' while preserving its
magnitude.  This was shown explicitly in a model wherein the spin
interaction for the quark is categorized into spin flip and non
spin-flip amplitudes during hadronization.  It was further shown that
the ratio of the amplitude magnitudes is given, at all energies and
angles, by the cosine of their phase difference; the consequences of
this relationship remain to be discovered.  While this model
hypothesis is somewhat heuristic, especially ignoring the initial
proton's spin, it may nevertheless serve as a starting point for
deeper considerations of this problem.


\begin{thebibliography}{}
%
%
\bibitem{bradford}
R. Bradford, R. A. Schumacher {\it et al.} (CLAS Collaboration), submitted to
Phys. Rev. C, arXiv:nucl-ex/0611034.


\bibitem{ambrozewicz}
P. Ambrozewicz, D. S. Carman, R. J. Feuerbach, M. D. Mestayer,
B. A. Raue, R. A. Schumacher , A. Tkabladze {\it et al.} (CLAS Collaboration), submitted to
Phys. Rev. C, arXiv:hep-ex/0611036.

\bibitem{mcnabb} J. W. C. McNabb, R. A. Schumacher, L. Todor (CLAS
Collaboration), {\it et al.},
Phys. Rev. C {\bf 69}, 042201(R) (2004).

\bibitem{barker} I. S. Barker, A. Donnachie, and J. K. Storrow,
  Nucl. Phys. {\bf B95}, 347 (1975).

\bibitem{bradforddsdo} R. Bradford, R.A. Schumacher, J. W. C. McNabb,
L. Todor, {\it et al.} (CLAS Collaboration), Phys. Rev. C {\bf 73}
035202 (2006).

\bibitem{merz}
This discussion uses the notation and general arguments found in 
Eugen Merzbacher, \textit{Quantum Mechanics, 2nd Ed.} (John Wiley and
Sons, New York, 1970) pp. 277-293.
\end{thebibliography}
\end{document}